%% file: manuscript.tex
\documentclass[%
 reprint,
superscriptaddress,
 amsmath,amssymb,
 aps,
]{revtex4-2}

\usepackage{graphicx}% Include figure files
\usepackage{dcolumn}% Align table columns on decimal point
\usepackage{bm}
\usepackage{siunitx}
 \usepackage{relsize}
\usepackage{placeins}
\usepackage{subfig}
\usepackage{caption}
\usepackage{times}
\usepackage{xcolor} 
\usepackage{float}
\usepackage{hyperref}
\hypersetup{
    colorlinks=true,
    linkcolor=blue,
citecolor=blue}
\makeatletter
\let\save@mathaccent\mathaccent
\newcommand*\if@single[3]{%
  \setbox0\hbox{${\mathaccent"0362{#1}}^H$}%
  \setbox2\hbox{${\mathaccent"0362{\kern0pt#1}}^H$}%
  \ifdim\ht0=\ht2 #3\else #2\fi
  }
\newcommand*\rel@kern[1]{\kern#1\dimexpr\macc@kerna}
\newcommand*\widebar[1]{\@ifnextchar^{{\wide@bar{#1}{0}}}{\wide@bar{#1}{1}}}
\newcommand*\wide@bar[2]{\if@single{#1}{\wide@bar@{#1}{#2}{1}}{\wide@bar@{#1}{#2}{2}}}
\newcommand*\wide@bar@[3]{%
  \begingroup
  \def\mathaccent##1##2{%
    \let\mathaccent\save@mathaccent
    \if#32 \let\macc@nucleus\first@char \fi
    \setbox\z@\hbox{$\macc@style{\macc@nucleus}_{}$}%
    \setbox\tw@\hbox{$\macc@style{\macc@nucleus}{}_{}$}%
    \dimen@\wd\tw@
    \advance\dimen@-\wd\z@
    \divide\dimen@ 3
    \@tempdima\wd\tw@
    \advance\@tempdima-\scriptspace
    \divide\@tempdima 10
    \advance\dimen@-\@tempdima
    \ifdim\dimen@>\z@ \dimen@0pt\fi
    \rel@kern{0.6}\kern-\dimen@
    \if#31
      \overline{\rel@kern{-0.6}\kern\dimen@\macc@nucleus\rel@kern{0.4}\kern\dimen@}%
      \advance\dimen@0.4\dimexpr\macc@kerna
      \let\final@kern#2%
      \ifdim\dimen@<\z@ \let\final@kern1\fi
      \if\final@kern1 \kern-\dimen@\fi
    \else
      \overline{\rel@kern{-0.6}\kern\dimen@#1}%
    \fi
  }%
  \macc@depth\@ne
  \let\math@bgroup\@empty \let\math@egroup\macc@set@skewchar
  \mathsurround\z@ \frozen@everymath{\mathgroup\macc@group\relax}%
  \macc@set@skewchar\relax
  \let\mathaccentV\macc@nested@a
  \if#31
    \macc@nested@a\relax111{#1}%
  \else
    \def\gobble@till@marker##1\endmarker{}%
    \futurelet\first@char\gobble@till@marker#1\endmarker
    \ifcat\noexpand\first@char A\else
      \def\first@char{}%
    \fi
    \macc@nested@a\relax111{\first@char}%
  \fi
  \endgroup
}
\makeatother

\newsavebox{\mybox}
\newlength{\mywidth}
\newlength{\myheight}
\newlength{\myline}
\newlength{\myoffset}
\newcommand{\mysqrt}[1]%
{\setlength{\myline}{.1ex}%
\addtolength{\myline}{.06pt}%
\setlength{\myoffset}{.9em}
\addtolength{\myoffset}{-2pt}
\savebox{\mybox}{$\displaystyle\sqrt{#1}$}%
\settoheight{\myheight}{\usebox{\mybox}}%
\addtolength{\myheight}{-.2ex}
\settowidth{\mywidth}{\usebox{\mybox}}%
\addtolength{\mywidth}{-\myoffset}%
 \rlap{\usebox{\mybox}}\hspace{\myoffset}{\raisebox{\myheight}{\rule{\mywidth}{\myline}}}}

\begin{document}

\preprint{APS/123-QED}

\title{$\alpha$-decay half-lives of superheavy nuclei with $Z=122-125$}

\author{Omar Nagib}%
\email{{\color{black} omar.khaled.nagib@gmail.com}}
\affiliation{Department of Physics, German University in Cairo, Cairo 11835, Egypt}

\date{\today}% It is always \today, today,

\begin{abstract}
For $\alpha$ decay half-life calculations in this work, the Coulomb and proximity potential model with a new semiempirical formula for diffuseness parameter developed in previous work [Phys. Rev. C 100, 024601 (2019)] is used. The present model in this work is compared with the generalized liquid-drop model (GLDM), universal decay law (UDL), and experimental half-lives in the region $Z=104-118$. Next, the predicted half-lives of 51 superheavy nuclei (SHN) with $Z=122-125$ by the present model are compared with those of GLDM, and UDL. The present model is revealed to be more accurate in reproducing experimental half-lives compared to GLDM and UDL. Moreover, it is found that the predictions of the present model and UDL are highly consistent while GLDM largely deviates from the other two. A study of the competition between $\alpha$ decay and spontaneous fission (SF) shows that $\alpha$ decay is the dominant mode. Among the studied SHN with $Z=122-125$,  ${}^{295-307}122$ and ${}^{314-320}125$ are identified as potential candidates whose half-lives are relatively long enough to be experimentally detected in the future through their $\alpha$-decay chains. The identified candidates are in good agreement with other recent work.  
\end{abstract}

\maketitle

%\tableofcontents

\section{\label{sec:level1}Introduction}

Shortly after its discovery by Rutherford and Geiger \cite{rutherford}, $\alpha$ decay was explained as a quantum tunneling process by Gamow in 1928 \cite{gamow}. To this day, the topic of $\alpha$ decay remains an important one in nuclear physics being the dominant mode of decay for superheavy nuclei (SHN) and a simple mode of decay compared to other modes (e.g., cluster decay and fission) \cite{alphamode,alphamode2}. Among many things, $\alpha$ decay reveals information about nuclear structure and stability and can help in identifying new superheavy elements \cite{detect,detect2,detect3,detect4}. Since the time of Gamow, many theoretical and empirical models with varying degrees of accuracy and sophistication were developed for the calculation and prediction of $\alpha$-decay half-life \cite{DFM1,DFM2,LDM,prox,proxies,model,model2,model3,model4,model5}.

In previous work with Abdul-latif \cite{paper}, the Coulomb and proximity potential model for the calculation of half-life was employed. A novel semiempirical formula for diffuseness was proposed and used to calculate and predict the half-lives of 218 SHN \cite {paper}. By incorporating the formula for diffuseness in the calculations, the model in the past work was able to reproduce experimental half-lives pretty accurately and better than a lot of popularly used models and semiempirical formulas (e.g., deformed Woods-Saxon model, UNIV, SemFIS, Viola-Seaborg, and Royer10 formulas). This work extends the previous work in two aspects. First, the improved model is compared with two more models, namely, the generalized liquid-drop model (GLDM) and the universal decay law (UDL). Second, half-lives of 51 superheavy nuclei in the region $Z=122-125$ are predicted. The outline of this paper is as follows: in Sec. \ref{sec:num2}, the theoretical model that will be used in the calculations of half-lives is described. In Sec. \ref{sec:num3}, the present model is compared with GLDM, UDL, and experimental half-lives in the region $Z=104-118$. Moreover, the present model is used to predict the half-lives of 51 SHN with $Z=122-125$ and compared with that of GLDM and UDL and their relative consistency is studied. The competition between $\alpha$ decay and spontaneous fission (SF) for $Z=122-125$ is studied. Finally, potential SHN candidates that can be detected in future experiments through their $\alpha$-decay chains are identified and compared with results from other work.  In Sec. \ref{sec:num4}, the main conclusions and a summary of the work are presented. 

\section{Theoretical framework} \label{sec:num2}

The effective potential for $\alpha$ decay consists of three parts, namely nuclear, Coulomb, and angular parts

\begin{equation} \label{eq:1}
V_{\text{eff}}(r)= V_N(r)+V_C(r)+V_l(r)
\end{equation}

where $r$ is the separation distance between the center of the daughter and $\alpha$ nuclei. The angular part will be neglected in the present study. For the nuclear potential $V_N(r)$, the proximity potential previously proposed by Zhang \textit{et al.} \cite{prox} is adopted:

\begin{equation} \label{eq:2}
V_N(s_0)=4 \pi b_{\text{eff}} \widebar{R} \gamma \phi(s_0)
\end{equation}

\begin{equation} \label{eq:3}
 \phi(s_0)=\dfrac{p_1}{1+\exp \Big(\dfrac{s_0+p_2}{p_3}\Big)}
\end{equation}

In the two-parameter Fermi distribution (2pF) of nuclear matter, $b_{\text{eff}}$ and the effective diffuseness $a_{\text{eff}}$ of the proximity potential are related as \cite{cluster}

\begin{equation} \label{eq:4}
b_{\text{eff}} =\dfrac{\pi}{\sqrt{3}} a_{\text{eff}}
\end{equation}

In many works, $b_{\text{eff}}$ is taken as a constant equal to 0.99-1 fm (equivalently $a_{\text{eff}}=0.54$ fm) \cite{constb1,constb2,constb3,constb4,consta1,consta2,consta3,consta4,probe}. In previous work with Abdul-latif, it was found that such an approximation yields unacceptable errors \cite{paper}. Hence, the semiempirical formula developed in the previous work will be employed since it was shown to be accurate in reproducing experimental half-lives \cite{paper}. The formula is given by

\begin{equation} \label{eq:5}
a_{\text{eff}}=-1.09535 + 0.012063Z + 0.0019759N
\end{equation}

$\gamma$ is given by

\begin{equation} \label{eq:6}
\gamma =0.9517 \Big [1-1.7826\Big(\dfrac{N-Z}{A}\Big)^2 \Big] \,   \si{\mega\eV \per  \femto \metre \squared}
\end{equation}

$ \widebar{R}=R_{\alpha}R_d/(R_{\alpha}+R_d)$ is the reduced radius of the $\alpha$-daughter system where the radius of each nucleus in terms of its mass number is given by  

\begin{equation} \label{eq:7}
R = 1.28 A^{1/3}+0.8A^{-1/3}-0.76
\end{equation}

$ \phi(s_0)$ is the universal function expressed in terms of the reduced separation distance $s_0$:

\begin{equation} \label{eq:8}
s_0=\dfrac{r-R_{\alpha}-R_d}{b_{\text{eff}}}
\end{equation}

The constants $p_1,p_2$, and $p_3$ appearing in Eq. \eqref{eq:3} are given by -7.65, 1.02, and 0.89, respectively. The Coulomb part $V_C(r)$ is given by

\begin{equation}\label{eq:9}
      V_C(r)=
      Z_{\alpha} Z_d e^2 \left\{ \begin{array}{ll}
            \dfrac{1}{r} & r\geq R_C \\
            \dfrac{1}{2 R_C}\Big[3-\Big(\dfrac{r}{R_C}\Big)^2\Big] & r< R_C
        \end{array} \right.
    \end{equation}

where  $Z_{\alpha}$ and $Z_d$ are the $\alpha$ and daughter charge numbers and $R_C=R_{\alpha} +R_{d}$. Using the Wentzel-Kramers-Brillouin (WKB) approximation, the half-life is given by

\begin{equation} \label{eq:10}
T_{1/2}=\dfrac{\pi \hbar \ln2}{P_o E_{\nu}}\Big[1+\exp(K) \Big]
\end{equation}

where $P_o,  E_{\nu}$, and $K$ are the preformation factor, zero-point vibration energy, and action integral, respectively. The action integral is given by

\begin{equation} \label{eq:11}
K= \dfrac{2}{\hbar} \int_{r_1}^{r_2} \sqrt{2 \mu (V_{\text{eff}}(r)-Q_{\alpha} )} dr
\end{equation}

where $Q_{\alpha}, \mu, r_1$ and $r_2$ are the decay energy, reduced mass, and first and second turning points, respectively. A classical description of $E_{\nu}$ shall be used since it was previously found \cite{freq} that classical and quantum mechanical (e.g., modified harmonic oscillator) approaches yield similar results, hence

\begin{equation} \label{eq:12}
E_{\nu}=\dfrac{\hbar \omega}{2}=\dfrac{\hbar \pi}{2 R_p}\sqrt{\dfrac{2 E_{\alpha}}{m_{\alpha}}}
\end{equation}

where $R_p,E_{\alpha}$, and $m_{\alpha}$ are the radius of the parent nucleus and kinetic energy and mass of the $\alpha$ particle, respectively. The kinetic energy of the $\alpha$ particle is given by \cite{extended}

\begin{equation} \label{eq:13}
E_{\alpha}=\dfrac{A_d}{A_p} Q_{\alpha}- \Big(6.53 Z_d^{7/5}-8 Z_d^{2/5} \Big) 10^{-5} \,  \si{\mega\eV}
\end{equation}

where $A_d$ and $A_p$ are the mass numbers of the daughter and the parent nuclei, respectively. Finally, the preformation factor in the present work is given by \cite{preform} 

\begin{multline} \label{eq:14}
 \log_{10} P_o = a+b(Z-Z_1)(Z_2-Z_1)+c(N-N_1)(N_2-N) \\
 +d A+e(Z-Z_1)(N-N_1)
\end{multline}

where $a,b,c,d$, and $e$ are given by 34.90593, 0.003011, 0.003717, -0.151216, and 0.006681, respectively. $Z_1,Z_2,N_1$, and $N_2$ are the magic numbers given by 82, 126, 152, and 184, respectively.

\section{Results and Discussion} \label{sec:num3}

%htp!

\begin{table}[H]
\caption{ Statistical comparison between models Prox, UDL, and GLDM. Half-life calculations for the three models are taken from Refs. \cite{paper,GLDMUDL}. } 
\label{table:1}
\begin{ruledtabular}
\begin{tabular}{c c c c c }
Model   &  $\sqrt{\overline {\delta^{2}}}$ & $\overline{|\delta|}$ & $\bar{\delta}$ & $|\Delta|_{\text{max}}$   \\ \hline
Prox & 0.41 & 0.36 & 0.064 & 0.926  \\
UDL & 0.641 & 0.435 & 0.141 & 2.671  \\
GLDM & 0.714 & 0.47 & -0.078 & 3.197  \\
\end{tabular}
\end{ruledtabular}
\end{table}

To assess the relative accuracy of the models, the half-life calculations of the present model (dubbed Prox), UDL, GLDM, and experimental half-lives in the region $Z=104-118$ were compared. For the present model, calculations of 68 SHN were available and taken from previous work \cite{paper}. For UDL and GLDM, half-life calculations of 40 SHN were available and taken from Ref. \cite{GLDMUDL}. Using this data set, statistical parameters that measure the accuracy of the models were computed. The statistical parameters include root-mean-square (rms) deviation, mean deviation, mean, and magnitude of maximum error $|\Delta|_{\text{max}}$. The rms deviation $\sqrt{\overline {\delta^{2}}}$ of each model from experimental half-lives is given by  

\begin{equation} \label{eq:15}
 \sqrt{\overline {\delta^{2}}}= \mysqrt{\dfrac{1}{M}\mathlarger {\sum \limits_{i}^{M} \Delta_i^2}}
\end{equation}

The mean deviation $\overline{|\delta|}$ is computed by 

\begin{equation} \label{eq:16}
\overline{|\delta|}=\dfrac{1}{M} \mathlarger{ \sum \limits_{i}^{M}| \Delta_i| }
\end{equation}

The mean is given by 

\begin{equation} \label{eq:17}
\bar\delta=\dfrac{1}{M} \mathlarger{ \sum \limits_{i}^{M} \Delta_i }
\end{equation}

where $\Delta=\log_{10}T_{\text{exp}}-\log_{10}T_{\text{calc}}$ is the deviation of calculated logarithm of half-life from the experimental one and

\begin{table*}[htp!]
\caption{ Model Prox $\log_{10} T_{\text{prox}}$  vs UDL $\log_{10} T_{\text{UDL}}$ vs GLDM $\log_{10} T_{\text{GLDM}}$. Values for $Q_{\alpha}$, $\log_{10}T_{\text{UDL}}$, and $\log_{10}T_{\text{GLDM}}$  are from Refs. \cite{GLDMUDL,UDL125}. } 
\label{table:2}
\begin{ruledtabular}
\begin{tabular}{c c c c c c}
$Z$   & $A$   & $Q_{\alpha}$      & $\log_{10}T_{\text{prox}}$    & $\log_{10}T_{\text{UDL}}$    & $\log_{10}T_{\text{GLDM}}$  \\ \hline
122 & 295 & 13.844 & -6.13846 & -6.298  & -5.095 \\
    & 296 & 13.705 & -5.98077 & -6.502  & -5.334 \\
    & 297 & 13.67  & -6.00675 & -5.997  & -4.923 \\
    & 298 & 13.65  & -6.05305 & -6.43   & -5.339 \\
    & 299 & 13.446 & -5.7426  & -5.594  & -4.626 \\
    & 300 & 13.115 & -5.16303 & -5.393  & -4.429 \\
    & 301 & 13.142 & -5.27817 & -5.013  & -4.16  \\
    & 302 & 13.234 & -5.51409 & -5.671  & -4.746 \\
    & 303 & 12.964 & -5.01958 & -4.678  & -3.92  \\
    & 304 & 12.742 & -4.59973 & -4.676  & -3.949 \\
    & 305 & 12.933 & -5.02622 & -4.647  & -4.047 \\
    & 306 & 13.068 & -5.32347 & -5.398  & -4.699 \\
    & 307 & 13.82  & -6.78193 & -6.455  & -5.801 \\
    & 308 & 14.653 & -8.24236 & -8.445  & -7.516 \\
    & 309 & 14.1   & -7.29515 & -7.014  & -6.356 \\
    & 310 & 13.377 & -5.95135 & -6.094  & -5.444 \\
123 & 300 & 14.703 & -7.73571 & -7.222  & -7.86  \\
    & 301 & 14.434 & -7.3549  & -7.216  & -6.036 \\
    & 302 & 14.519 & -7.5682  & -6.932  & -7.524 \\
    & 303 & 14.588 & -7.74547 & -7.525  & -6.366 \\
    & 304 & 14.393 & -7.46473 & -6.74   & -7.369 \\
    & 305 & 14.28  & -7.31324 & -7.004  & -6.081 \\
    & 306 & 14.313 & -7.40856 & -6.627  & -7.198 \\
    & 307 & 14.33  & -7.46828 & -7.128  & -6.263 \\
    & 308 & 14.89  & -8.43457 & -7.682  & -8.055 \\
    & 309 & 15.382 & -9.23025 & -8.977  & -7.951 \\
    & 310 & 14.613 & -7.99778 & -7.233  & -7.586 \\
    & 311 & 13.736 & -6.44294 & -6.08   & -5.468 \\
124 & 301 & 15.012 & -7.99617 & -7.963  & -6.557 \\
    & 302 & 14.811 & -7.74724 & -8.085  & -6.638 \\
    & 303 & 14.889 & -7.95095 & -7.784  & -6.469 \\
    & 304 & 15.016 & -8.22584 & -8.475  & -7.146 \\
    & 305 & 14.902 & -8.09863 & -7.842  & -6.708 \\
    & 306 & 14.819 & -8.01316 & -8.168  & -6.939 \\
    & 307 & 14.807 & -8.03753 & -7.709  & -6.673 \\
    & 308 & 14.794 & -8.05269 & -8.158  & -6.984 \\
    & 309 & 15.331 & -8.95    & -8.635  & -7.528 \\
    & 310 & 15.54  & -9.29557 & -9.449  & -8.155 \\
    & 311 & 14.773 & -8.08294 & -7.717  & -6.799 \\
    & 312 & 13.891 & -6.53971 & -6.568  & -5.667 \\
125 & 310 & 15.938 & -9.6879  & -9.916  &        \\
    & 311 & 15.873 & -9.61948 & -9.613  &        \\
    & 312 & 15.11  & -8.45279 & -8.309  &        \\
    & 313 & 14.319 & -7.12356 & -6.846  &        \\
    & 314 & 14.023 & -6.59493 & -6.234  &        \\
    & 315 & 13.811 & -6.19899 & -5.862  &        \\
    & 316 & 13.761 & -6.09571 & -5.774  &        \\
    & 317 & 13.717 & -5.9958  & -5.698  &        \\
    & 318 & 13.374 & -5.30447 & -4.989  &        \\
    & 319 & 12.919 & -4.34303 & -3.996  &        \\
    & 320 & 12.535 & -3.4766  & -3.1211 & 
\end{tabular}
\end{ruledtabular}
\end{table*}

%\FloatBarrier

\begin{figure*}
 \centering
\subfloat{%
  \includegraphics[width=.5\textwidth]{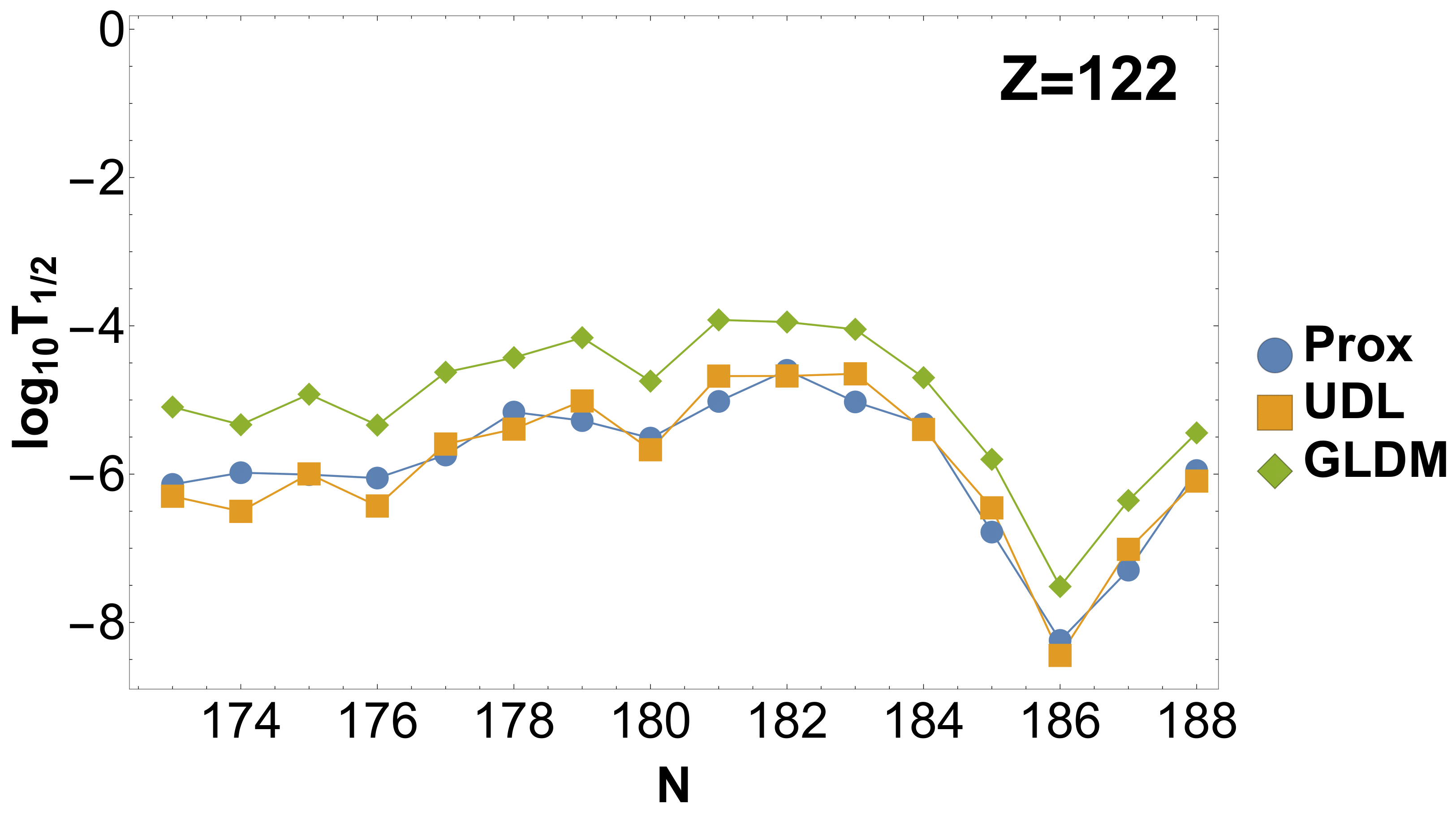}%
}
\subfloat{%
  \includegraphics[width=.5\textwidth]{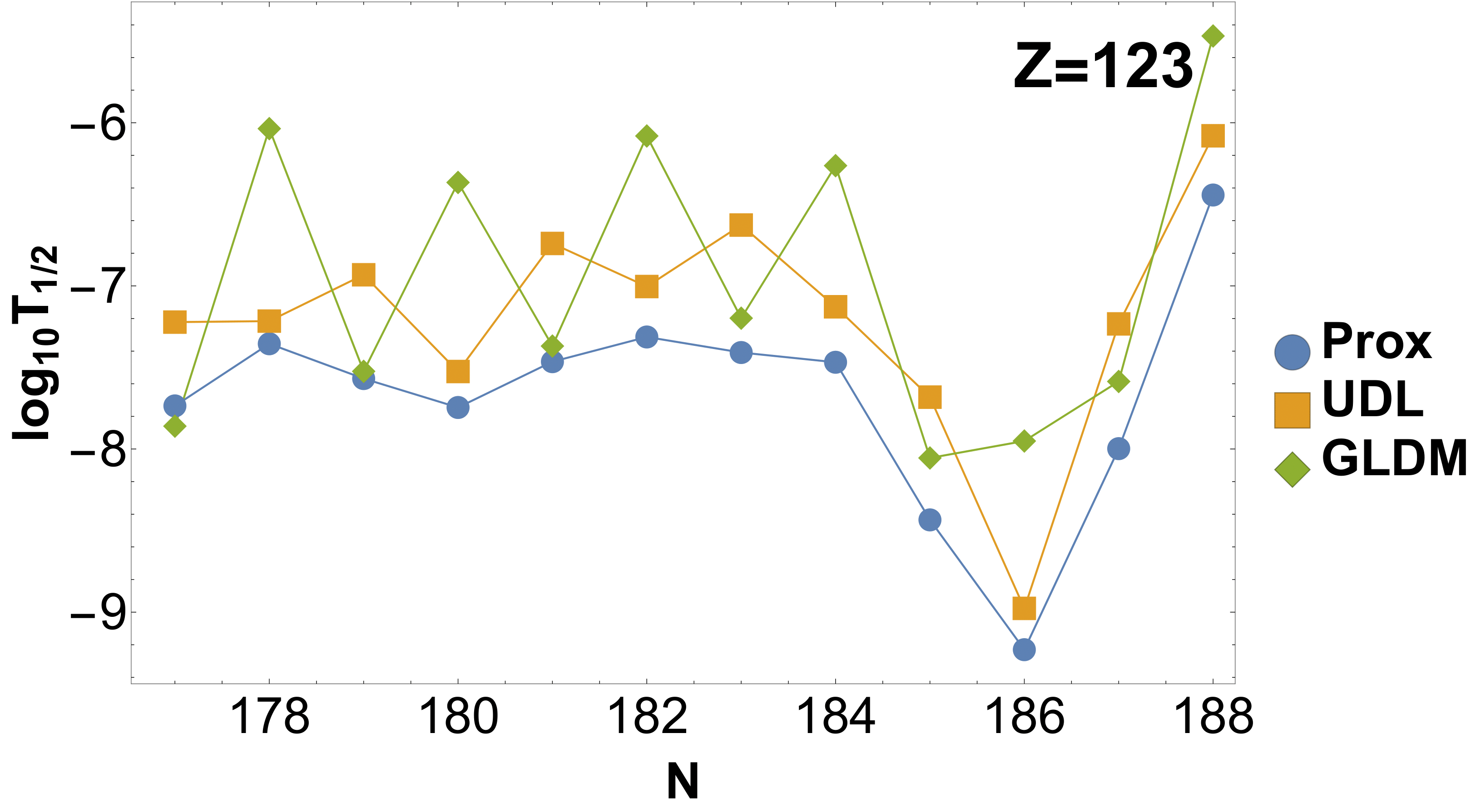}%
}\hfil
\subfloat{%
  \includegraphics[width=.5\textwidth]{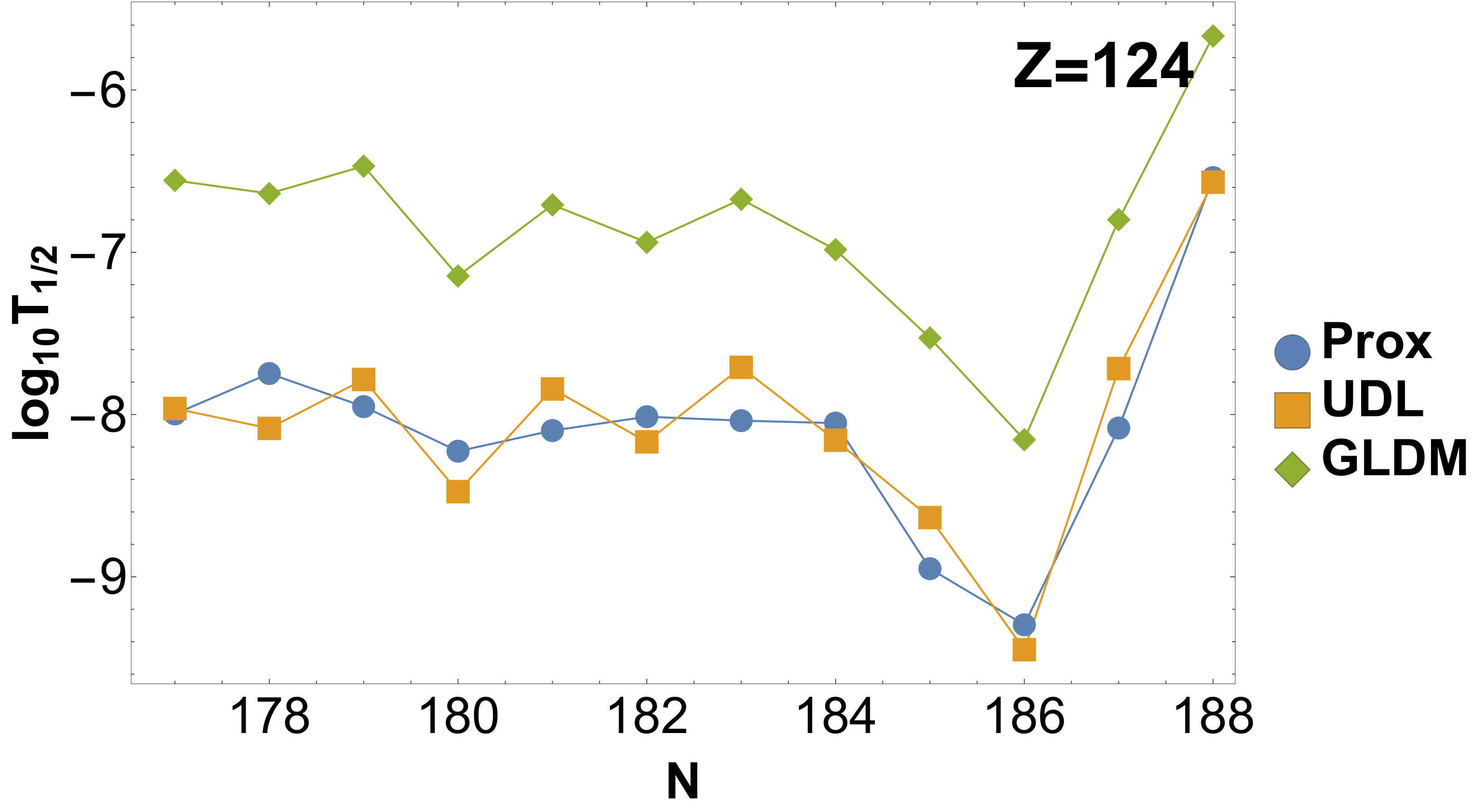}%
}
\subfloat{%
  \includegraphics[width=.5\textwidth]{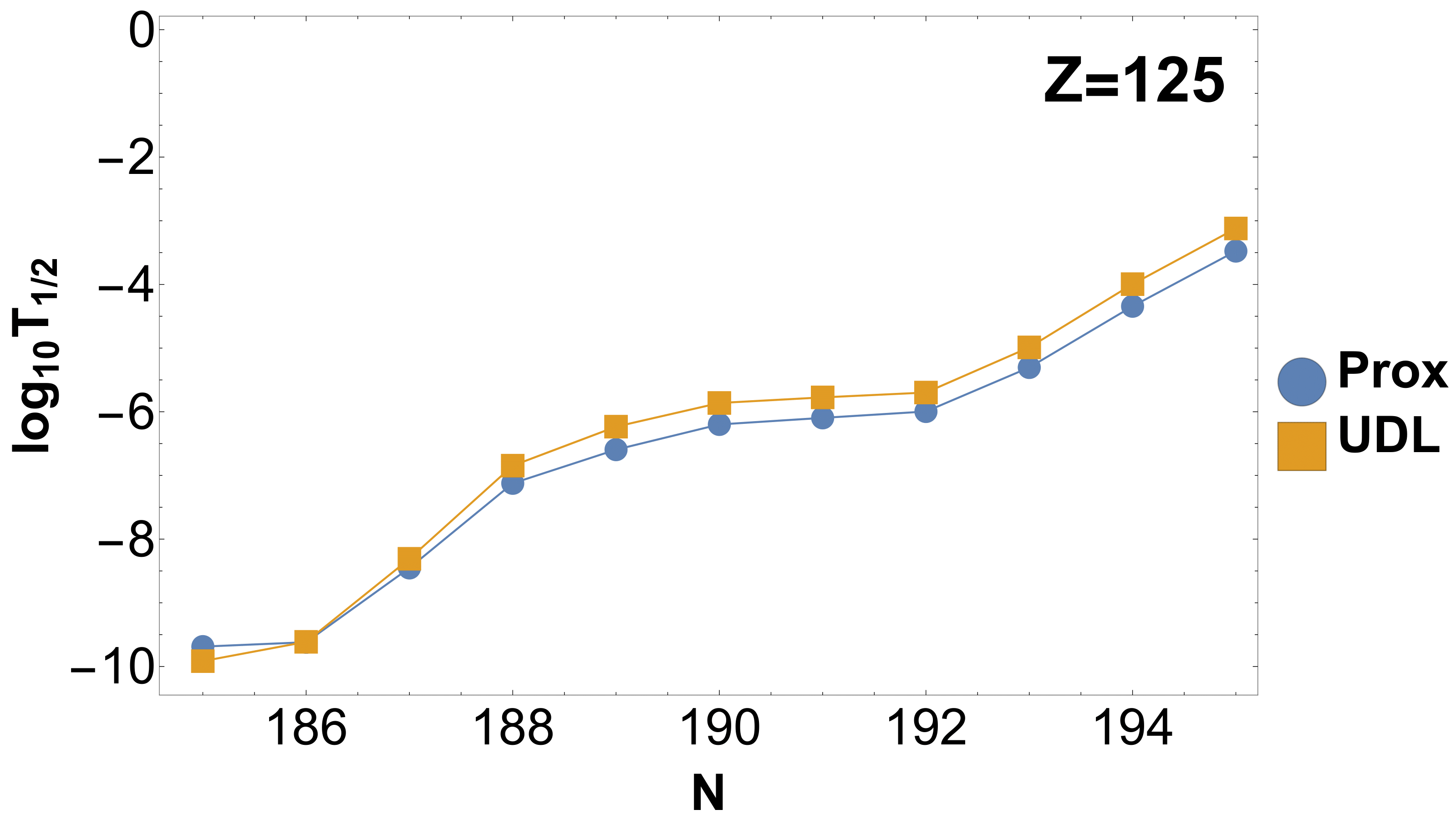}%
}      \caption{Plot of the logarithm of half-life (seconds) for the models Prox, UDL, and GLDM vs parent neutron number $N$ for $Z=122-125$. }
        \label{fig:Z}
\end{figure*}

 $M$ is the number of SHN under study. The results of these calculations are shown in Table \ref{table:1}. Looking at rms and mean deviations, one concludes that the present model is the most accurate in reproducing experimental half-lives, followed by UDL then GLDM. Looking at the maximum error, one sees that the deviation of GLDM from true half-life values can be as large as 3.19 (i.e., three orders of magnitude off) while the maximum error of UDL is 2.671 (i.e., off by a factor of 470) and the Prox model's maximum deviation from true half-life values does not exceed 0.926 (i.e., approximately off by a factor of 8). This statistical analysis concludes that the present model is the best one overall. Next, the half-lives of 51 SHN with $Z=122,123,124$, and $125$ were predicted using the present model and compared to those of UDL and GLDM. The values of $Q_{\alpha}$ and half-life predictions of UDL and GLDM were taken from Refs. \cite{GLDMUDL,UDL125}. The predicted half-lives using the three models are shown in Table \ref{table:2} with the values used for $Q_{\alpha}$. One notes that the calculations for $Z=125$ were not available for GLDM. Fig. \ref{fig:Z} displays plots of the predicted logarithm of half-life for the Prox model, UDL, and GLDM vs parent neutron number $N$ for the sake of visual comparison. From the figure, one makes two observations. The first observation to make is that the predictions of Prox model and UDL are very similar to each other for all $Z$ while GLDM deviates from the other two. Moreover, one finds the largest discrepancy between the three models for $Z=123$, where one also finds GLDM to exhibit a peculiar zig-zag pattern that the other two models do not show. To measure the consistency of predicted half-lives and how close the predictions are of one model with another, the rms deviation of each model with respect to the other was computed. For instance, to measure consistency of the Prox model with UDL, one defines $\Delta=\log_{10}T_{\text{UDL}}-\log_{10}T_{\text{prox}}$ and, using Eq. \eqref{eq:15} and the data in Table \ref{table:2}, one computes the rms deviation of the Prox model with respect to UDL. The same process is applied for the UDL/GLDM and GLDM/Prox cases. The results of these calculations are shown in Table \ref{table:3}. It is interesting to find that Prox and UDL are highly consistent and similar in their half-life predictions, with rms error of 0.342. However, one finds that the predictions of GLDM deviate significantly from the predictions made by UDL and Prox with rms deviations of 0.969 and 1.006, respectively. 

It is worth investigating the reasons behind the deviations of predictions of GLDM from that of the other two models in the region $Z=122-125$ since it is a somewhat unexpected result given that the three models are relatively consistent with each other and with the experimental half-lives in the region $Z=104-118$.  A potential cause for this behavior was identified. This has to do with the fact that, in GLDM, the diffuseness parameter $b_{\text{eff}}$ is set to a constant equal to 0.99 fm for all SHN \cite{GLDMroyer, GLDMUDL, MGLDM, GLDMroyer2}. The GLDM potential is the sum of volume, surface, Coulomb, and proximity potentials. The diffuseness parameter $b_{\text{eff}}$ enters half-life calculations in GLDM through the proximity energy $V_N(r)$ given by Eq. \eqref{eq:2}. It was shown in previous work with Abdul-latif and in the work of Dehghani \textit{et al.} that the half-life of a given nucleus decreases with increasing diffuseness parameter (given that all other parameters like decay energy are held fixed) \cite{roleconst, paper}. When the proximity potential is used, it was found that half-life depends more strongly on $b_{\text{eff}}$ for larger values of $b_{\text{eff}}$ (e.g., a change of 0.1 fm around $b_{\text{eff}}=1.1$ fm will induce more change in half-life compared to a change of 0.1 fm around $b_{\text{eff}}=0.4$ fm). Moreover, it was demonstrated that the common approximation $b_{\text{eff}}=0.99$ fm for all SHN is not acceptable, especially in the region $b_{\text{eff}} \ge 0.99$ fm, since an error on the order of 0.1 fm can yield an error in the logarithm of half-life as large as 0.7 \cite{paper}. Note that according to the formula for diffuseness, Eq. \eqref{eq:5}, $1.06 \le b_{\text{eff}} \le 1.17 $ fm for SHN in the region $Z=122-125$. Therefore, the approximation $b_{\text{eff}}=0.99$ fm employed in GLDM always underestimates the true value of diffuseness, which results in its half-life predictions in $Z=122-125$ being systematically higher than the other two models (see Fig. \ref{fig:Z}) since decreasing diffuseness increases half-life. The remedy to this problem is to use an accurate formula for diffuseness that is dependent on $Z$ and $N$, like the one used in the present work.

$\alpha$-decay half-life is among the quantities that are measured experimentally, and hence its accurate theoretical prediction is extremely important for discovering new SHN. Another crucial signature that identifies a nucleus is its $\alpha$-decay chain when it undergoes several $\alpha$-decays followed by SF. Therefore, the competition between $\alpha$-decay and SF in the region $Z=122-125$ shall be studied next. SHN which have relatively short $\alpha$-decay half-life compared to SF half-life will survive fission and hence can be synthesized in the laboratory and be detected by their $\alpha$-decay chains. Therefore, SHN in $Z=122-125$ that survive fission and whose $\alpha$-decay half-lives are long enough ($\ge 100$ ns) to be detected in future experiments shall be identified.

To this end, one defines the branching ratio $b$ of $\alpha$ decay with respect to spontaneous fission as

\begin{equation} \label{eq:18}
b=\dfrac{\lambda_{\alpha}}{\lambda_{\text{SF}}}=\dfrac{T_{\text{SF}}}{T_{\alpha}}
\end{equation}

where $\lambda_{\alpha}$ and $T_{\alpha}$ denote the $\alpha$-decay constant and half-life while $\lambda_{\text{SF}}$ and $T_{\text{SF}}$ denote those of SF. $\log_{10} b > 0$ implies that $\alpha$ decay is dominant and the particular SHN will survive fission and decay through $\alpha$-decay chains while SHN with $\log _{10} b < 0$ will not survive fission.

With the aid of the predicted $\alpha$-decay half-lives of the present model and that of UDL and the data for SF half-life $T_{\text{SF}}$ taken from Refs. \cite{GLDMUDL,UDL125}, Table \ref{table:4} was compiled where $\log_{10} T_{\text{SF}}$, $\log_{10} b_{\text{prox}}$, and $\log_{10} b_{\text{UDL}}$ are displayed. One finds that $\alpha$ decay is the dominant mode, as evidenced by the fact that $\log_{10} b_{\text{prox}} \le 0$ for only four SHN out of 51. UDL gives the same result with $\log_{10} b_{\text{UDL}} \le 0$ for the same four SHN. Therefore, one concludes that almost all of SHN in $Z=122-125$ will survive fission. The four SHN that will not survive fission are ${}^{309-310}122$, ${}^{311}123$, and ${}^{312}124$. One further notices that all SHN with $Z=125$ will survive fission. SHN with comparable SF and $\alpha$-decay half-lives are ${}^{308-309}122$ and ${}^{310-311}124$.

\begin{table}[H]
\caption{ rms deviations to measure consistency of one model with respect to the other.} 
\label{table:3}
\begin{ruledtabular}
\begin{tabular}{c c }
Model  &  $\sqrt{\widebar {\delta^{2}}}$  \\ \hline
Prox/UDL & 0.347  \\
GLDM/UDL & 0.969  \\
Prox/GLDM & 1.006  \\
\end{tabular}
\end{ruledtabular}
\end{table}

Using the calculations displayed in Table \ref{table:2} and \ref{table:4}, a search was carried out for SHN which both survive fission and whose predicted half-lives are relatively long enough ($\ge 100$ ns) to be detected in future experiments. By stipulating these requirements, it was found that ${}^{295-307}122$ and ${}^{314-320}125$ are potential candidates to be experimentally detected and identified via their decay chains. All other SHN either do not survive fission or have very short half-lives to be detected. Note that the isotopes under study in this paper are $295 \le A \le 310$ for $Z=122$, $300 \le A \le 311$ for $Z=123$, $301 \le A \le 312$ for $Z=124$, and $310 \le A \le 320$ for $Z=125$. Therefore, there could be other potential candidates that are experimentally detectable for $Z=122-125$ for larger or smaller $A$ which were not investigated here. 

Compared to a series of recent work done Santhosh \textit{et al.} \cite{UDL125, 99Z129, KPSZ123, Feasibility}, one finds that there is a good agreement with the experimentally detectable SHN for $Z=122$, where they concluded that ${}^{298-307}122$ are potential candidates. There is also a good agreement for $Z=125$, where they predict ${}^{310-320}125$ as potential candidates, thus agreeing with the present work on the detectability of ${}^{314-320}125$ and disagreeing on ${}^{310-313}125$. Again, there is a relatively good agreement for $Z=124$ where the present work excludes all SHN with $301 \le A \le 312$, while their model also excludes all SHN except for the four SHN ${}^{305-308}124$. The biggest disagreement is for $Z=123$, where this work excludes all SHN with $300 \le A \le 311$, while the other work posits that ${}^{300-307}123$ are detectable. All in all, the level of agreement between the present paper and the others is satisfactory given the different decay energies, SF half-lives, and theoretical models employed to determine possible candidates.

%\begin{table}[htp!]%The best place to locate the table environment is directly after its first reference in text
%\caption{\label{tab:table1}%
%A table that fits into a single column of a two-column layout. 
%Note that REV\TeX~4 adjusts the intercolumn spacing so that the table fills the
%entire width of the column. Table captions are numbered
%automatically. 
%This table illustrates left-, center-, decimal- and right-aligned columns,
%along with the use of the \texttt{ruledtabular} environment which sets the 
%Scotch (double) rules above and below the alignment, per APS style.
%}
%\begin{ruledtabular}
%\begin{tabular}{lcdr}
%\textrm{Left\footnote{Note a.}}&
%\textrm{Centered\footnote{Note b.}}&
%\multicolumn{1}{c}{\textrm{Decimal}}&
%\textrm{Right}\\
%\colrule
%1 & 2 & 3.001 & 4\\
%10 & 20 & 30 & 40\\
%100 & 200 & 300.0 & 400\\
%\end{tabular}
%\end{ruledtabular}
%\end{table}

\begin{table*}[t]
\caption{ Logarithm of SF half-life $\log_{10} T_{\text{SF}}$ taken from Refs. \cite{GLDMUDL,UDL125} with $\log_{10} b_{\text{prox}}$ and $\log_{10} b_{\text{UDL}}$ denoting the logarithm of branching ratio where predicted $\alpha$-decay half-lives of models Prox and UDL were used.  } 
\label{table:4}
\begin{ruledtabular}
\begin{tabular}{c c c c c}
$Z$& $A$ & $\log_{10} T_{\text{SF}}$   & $\log_{10} b_{\text{prox}}$   & $\log_{10} b_{\text{UDL}}$  \\ \hline
122 & 295 & 7.301    & 13.43946  & 13.599   \\
    & 296 & 4.016    & 9.99677   & 10.518   \\
    & 297 & 6.772    & 12.77875  & 12.769   \\
    & 298 & 3.677    & 9.73005   & 10.107   \\
    & 299 & 6.587    & 12.3296   & 12.181   \\
    & 300 & 5.302    & 10.46503  & 10.695   \\
    & 301 & 8.806    & 14.08417  & 13.819   \\
    & 302 & 4.937    & 10.45109  & 10.608   \\
    & 303 & 5.992    & 11.01158  & 10.67    \\
    & 304 & 1.195    & 5.79473   & 5.871    \\
    & 305 & 2.875    & 7.90122   & 7.522    \\
    & 306 & -2.21    & 3.11347   & 3.188    \\
    & 307 & -3.116   & 3.66593   & 3.339    \\
    & 308 & -8.084   & 0.15836   & 0.361    \\
    & 309 & -7.892   & -0.59685  & -0.878   \\
    & 310 & -13.149  & -7.19765  & -7.055   \\
123 & 300 & 10.421   & 18.15671  & 17.643   \\
    & 301 & 9.448    & 16.8029   & 16.664   \\
    & 302 & 13.333   & 20.9012   & 20.265   \\
    & 303 & 11.063   & 18.80847  & 18.588   \\
    & 304 & 8.852    & 16.31673  & 15.592   \\
    & 305 & 4.034    & 11.34724  & 11.038   \\
    & 306 & 5.661    & 13.06956  & 12.288   \\
    & 307 & 0.388    & 7.85628   & 7.516    \\
    & 308 & -0.4     & 8.03457   & 7.282    \\
    & 309 & -5.249   & 3.98125   & 3.728    \\
    & 310 & -4.498   & 3.49978   & 2.735    \\
    & 311 & -9.296   & -2.85306  & -3.216   \\
124 & 301 & 7.676    & 15.67217  & 15.639   \\
    & 302 & 8.836    & 16.58324  & 16.921   \\
    & 303 & 12.76    & 20.71095  & 20.544   \\
    & 304 & 10.464   & 18.68984  & 18.939   \\
    & 305 & 14.209   & 22.30763  & 22.051   \\
    & 306 & 0.176    & 8.18916   & 8.344    \\
    & 307 & 1.746    & 9.78353   & 9.455    \\
    & 308 & -3.784   & 4.26869   & 4.374    \\
    & 309 & -3.748   & 5.202     & 4.887    \\
    & 310 & -8.684   & 0.61157   & 0.765    \\
    & 311 & -7.242   & 0.84094   & 0.475    \\
    & 312 & -11.922  & -5.38229  & -5.354   \\
125 & 310 & 22.64246 & 32.33036  & 32.55846 \\
    & 311 & 21.40278 & 31.02226  & 31.01578 \\
    & 312 & 20.48101 & 28.9338   & 28.79001 \\
    & 313 & 19.11059 & 26.23415  & 25.95659 \\
    & 314 & 17.99047 & 24.5854   & 24.22447 \\
    & 315 & 16.39863 & 22.59762  & 22.26063 \\
    & 316 & 14.89048 & 20.98619  & 20.66448 \\
    & 317 & 12.95885 & 18.95465  & 18.65685 \\
    & 318 & 11.14114 & 16.44561  & 16.13014 \\
    & 319 & 5.625929 & 9.968959  & 9.621929 \\
    & 320 & 3.8892   & 7.3658    & 7.0103   \\  
\end{tabular}
\end{ruledtabular}
\end{table*}

\section{summary and conclusions} \label{sec:num4}

In this work, the Coulomb and proximity potential model with a new semiempirical formula for diffuseness parameter was used for half-life calculations.  The half-life calculations of the present model, UDL, GLDM, and experimental half-lives in the region $Z=104-118$ were compared. Various statistical parameters show that the present model is the most accurate in reproducing experimental half-lives followed by UDL then GLDM. The predicted half-lives of 51 SHN in the $Z=122-125$ region by the present model were compared with those of UDL and GLDM. The predictions of the present model and UDL are highly consistent while GLDM largely deviates from the two. The deviation of GLDM in the region $Z=122-125$ is potentially caused by the approximation $b_{\text{eff}}=0.99$ fm which leads to systematic deviation and overestimation of true half-life values. The study of competition between $\alpha$-decay and SF for $Z=122-125$ concludes that $\alpha$-decay is the dominant mode. SHN that will not survive fission and those with comparable SF and $\alpha$-decay half-lives were also identified. Moreover, using the present calculations, a systematic search was carried for SHN that survive fission and whose predicted half-lives are within experimental limits so that they can be detected and identified via their $\alpha$-decay chains. SHN ${}^{295-307}122$ and ${}^{314-320}125$ were identified as candidates to be detected. The identified candidates are in good agreement with other recent work. One hopes that the present half-life calculations and identified candidates will guide future experiments in detecting new SHN.

\input{manuscript.bbl}
\end{document}
%
% ****** End of file apssamp.tex ******

%% file: manuscript.bbl
\providecommand{\noopsort}[1]{}\providecommand{\singleletter}[1]{#1}%